# Review and Outlook of Solar-Energetic-Particle Measurements on Multispacecraft Missions

*Donald V. Reames*

Institute for Physical Science and Technology, University of Maryland, College Park, MD, USA

dvreames@gmail.com

**Abstract** The earliest evidence on spatial distributions of solar energetic particles (SEPs) compared events from many different source longitudes on the Sun, but the early *Pioneers* provided the first evidence of the large areas of equal SEP intensities across the magnetically-confined "reservoirs" late in the events. More-detailed measurements of the importance of self-generated waves and trapping structures around the shock waves that accelerate SEPs were obtained from the *Helios* mission plus IMP 8, especially during the year when the two *Voyager* spacecraft also happened by. The extent of the dozen widest SEP events in a solar cycle, that effectively wrap around the Sun, was revealed by the widely separated STEREO spacecraft with three-point intensities fit to Gaussians. Element abundances of the broadest SEP events favor average coronal element abundances with little evidence of heavy-element-enhanced "impulsive suprathermal" ions that often dominate the seed population of the shocks, even in extremely energetic local events. However, it is hard to define a distribution with two or three points. Advancing the physics of SEPs may require a return to the closer spacing of the *Helios* era with coverage mapped by a half-dozen spacecraft to help disentangle the distribution of the SEPs from the underlying structure of the magnetic field and the accelerating shock.

## 1      Introduction

The spatial distribution of solar energetic particles (SEPs), and its variation with time, particle species, and energy, is fundamental to an understanding of the physics of particle acceleration and transport in the heliosphere. How much of the variation we see at a single spacecraft is a true time variation and how much is spatial variation being convected past? Is the SEP source itself broadly extended in space or do SEPs somehow diffuse out of a limited source? Does a shock source sample different abundances of seed ions from different places or, for example, does Fe simply scatter less than O, to produce early enhancements and later suppressions in Fe/O? Multispacecraft comparisons can be a key to distinguishing the physical effects dependent upon space and time.

### 1.1     SEP History and Context

Multispacecraft measurements, and the perceived need for them, generally followed the study of solar energetic particles (SEPs) on or near Earth. The SEP events observed first (Forbush 1946) were rare, large, energetic "ground-level enhancements" (GLEs) where GeV protons produce a nuclear cascade through the atmosphere to ground level that enhances the continuous signal produced similarly by the galactic cosmic rays (GCRs). Since solar flares were found to accompany these early SEP events, flares were considered a possible source. However, the spatial span of these flares, stretching from the east on the Sun to behind the western limb, raised a significant problem: how could the SEPs cross magnetic field lines radiating out from the Sun to find their way to Earth?

Meanwhile, solar radio astronomers, also using ground-based instruments, had identified different sources triggered by energetic solar electrons. Radio emission, excited at the local plasma frequency depends upon the square root of the local electron density which decreases with distance from the Sun. Wild et al. (1963) described radio type III bursts where frequencies decreased rapidly, excited by 10 – 100 keV electrons that streamed out from a source near the Sun. There is also type II bursts where the source moved at the slower speed of a ~1000 km s$^{-1}$ shock wave. Wild et al. (1963) suggested two types of SEP sources: point sources of mostly electrons near the Sun and fast shock waves that could accelerate energetic protons, like those that produce GLEs. Even though shock acceleration was well known in other contexts, like the supernova sources of GCRs, Wild et al. (1963) were far ahead of their time in solar physics. Twenty years later, the clear 96% association of large SEP events with shocks driven by fast, wide coronal mass ejections (CMEs) was established by Kahler et al. (1984). Yet ten years after that, Gosling (1993, 1994) still needed to point out the error of the "Solar Flare Myth".

Parker (1965) explained particle transport in terms of pitch-angle scattering as SEPs followed magnetic field lines out from the Sun. Diffusion theory is an important tool when there is actually a physical mechanism, like pitch-angle scattering, to produce the random walk. There is also a random walk of the magnetic field footpoints (Jokipii and Parker 1969; Li and Bian 2023) prior to events, that can produce an effective random walk perpendicular to the mean magnetic field, but it is independent of SEP-event parameters and is completely inadequate to explain a huge spread of SEPs far from a presumed source longitude near a flare. Other schemes such as the "birdcage" model (Newkirk and Wenzel 1978) were also invented to spread SEPs from a flare source (see review, Sec. 2.3 in Reames 2021a). Reinhard and Wibberenz (1974) envisioned a mysterious "fast propagation region" extending 60º from the flare to spread the SEPs prior to their slower interplanetary journey. Could this region actually match the surface of a shock? After decades of resistance, a flare source has mainly been abandoned for the largest SEP events that are now generally attributed to spatially extensive CME-driven shock waves (Mason et al 1984; Gosling 1993, Reames 1995b, 1999, 2013, 2021b; Zank et al. 2000, 2007; Lee et al. 2012; Desai and Giacalone 2016), especially for GLEs (Tylka and Dietrich 2009; Gopalswamy et al. 2012, 2013; Mewaldt et al. 2012; Raukunen et al. 2018). Observations (e.g. Kahler et al. 1984) did replace the "fast propagation region" with the surface of a CME-driven shock wave that actually accelerates the particles, beginning at 2 – 3 solar radii (Tylka et al 2003; Reames 2009a, 2009b; Cliver et al. 2004), and continuing far out into the heliosphere. We will soon see from STEREO observations (e.g. Figure 5) that shocks easily wrap around the Sun expanding widely across magnetic field lines where SEPs alone cannot go.

As the element and isotope abundances in SEPs began to be measured they would present new evidence for two different physical sources of SEPs. The earliest measurements, during large SEP events with nuclear emulsions on sounding rockets, extended element abundances up to S (Fichtel and Guss 1961) then to Fe (Bertsch et al. 1969). Later studies showed that average SEP element abundances in large events were a measure of coronal abundances that differed from photospheric abundances as a simple function of the elements' first ionization potential (FIP; Meyer 1985; Reames 1995, 2014, 2021a, b). The FIP-dependence of SEPs differs fundamentally from that of the solar wind (Reames 2018a; Laming et al. 2019), probing the physics of formation of the corona itself.

However, early measurements in space soon identified a completely new type of event, distinguished by extremely high abundances of $^3$He in some events, such as $^3$He/$^4$He =1.52±0.1 (e.g. Serlemitsos and Balasubrahmanyan 1975), vs. a solar value of ~5 × 10$^{-4}$. Production of $^3$He from $^4$He by nuclear fragmentation was ruled out by the lack of $^2$H and by lack of Li, Be, and B fragments from C and O, e.g. Be/O and B/O < 2 × 10$^{-4}$ (McGuire et al., 1979; Cook et al., 1984). The huge enhancements of

³He were produced by new physics, involving a wave-particle resonance (e.g. Fisk 1978; Temerin and Roth 1993) with complex spectra (Mason 2007; Liu et al. 2004, 2006). The ³He-rich events were associated with the beamed non-relativistic electrons (Reames et al 1985) and their type III radio bursts (Reames and Stone 1986) studied by Wild et al. (1965). Element abundances of these "impulsive" ³He-rich events were also different from those of the large "gradual" shock-associated events (Mason et al. 1986; Reames 1988); here the enhancements increased as a power of the ion mass-to-charge ratio *A/Q* (Reames et al. 1994), which was especially clear when it became possible to measure elements above Fe with element groups resolved as high as Au and Pb (Reames 2000; Mason et al. 2004; Reames and Ng 2004). Average enhancements varied as $(A/Q)^{3.6}$ with $Q$ value determined at a temperature of ~3 MK (Reames et al. 2014a, b). In fact, this power law can be used in a best-fit method to determine source plasma temperatures (Reames 2016, 2018b). Impulsive SEP events have been associated with magnetic reconnection (Drake et al. 2009) on open field lines in solar jets (Bučík 2020) which also eject CMEs that sometimes drive shocks (Kahler et al 2001; Nitta et al. 2006, 2015; Wang et al. 2006; Bučík et al. 2018a, 2018b, 2021) fast enough to reaccelerate the enhanced ³He and heavy ions along with ambient protons and ions (Reames 2019, 2022b). Derived abundances from γ-ray lines suggest that flares involve similar physics (Mandzhavidze et al. 1999; Murphy et al. 1991, 2016), but those accelerated particles are trapped on loops, losing their energy to γ-rays, electron bremstrahlung, and hot, bright plasma. The opening magnetic reconnections that drive jets also must close neighboring fields to form flares (see e.g. Figure 4 in Reames 2021b).

Shock waves can also accelerate residual suprathermal impulsive ions that can pool to provide a seed population (Mason et al. 1999, Tylka et al. 2001, 2005; Desai et al. 2003; Tylka and Lee 2006). The combination of two fundamental physical acceleration mechanisms, magnetic reconnection and shock acceleration, and two distinctive element abundance patterns, led Reames (2020, 2022b) to suggest four distinguishable SEP-event abundance pathways:
  SEP1: "Pure" impulsive magnetic reconnection in solar jets with no fast shock.
  SEP2: Jets with fast, narrow CMEs driving shocks that reaccelerate SEP1 ions plus ambient coronal plasma. Pre-enhanced SEP1 ions dominate high $Z$, ambient protons dominate low $Z$.
  SEP3: Fast, wide CME-driven shocks accelerate SEP1 residue from active-region pools from many jets, plus ambient plasma. Again the SEP1 seed ions dominate high $Z$.
  SEP4: Fast, wide CME-driven shocks accelerate ions where ambient plasma completely dominates.

Persistent pools of SEP1-residual seed ions available for reacceleration in SEP3 events have now been widely observed and reported (Richardson et al. 1990; Desai et al. 2003; Bučík et al. 2014, 2015; Chen et al. 2015; Reames 2022a) and may be fed by numerous impulsive events too small to be distinguished individually.

Large gradual SEP events are frequently accompanied initially by type-III bursts. In principle, these impulsive jets could inject a SEP1 seed population. However, any such injection of SEP1 ions seems to be swamped by coronal seed ions in the many large SEP4 events. The extremely abundant type-III electrons can be distinguished from shock-accelerated electrons by their proximity to the event source; the latter emerge only in poorly-connected events (Cliver and Ling 2007).

The spatial extent of these features and the underlying physics is of considerable interest. Can we find spatial differences in the seed populations sampled by shocks? Unfortunately, each point on a shock moving radially across Parker-spiral fields will have spread particles over 50⁰ - 60⁰ by the time it reaches 1 AU (Reames 2022a). This tends to blur any initial spatial variations in abundances.

## 2    Near-Earth Observations

Before we could monitor a single SEP event at multiple locations we could study multiple events at Earth from different source longitudes on the Sun. Using data from IMP 4, 5, 7, 8 and ISEE 3 over nearly 20 years, Cane et al. (1988) ordered the 20 MeV proton profiles of 235 large SEP events as a function of solar source longitude. This study gave the typical extent and time profiles of an "average" SEP event, sliced as a function of longitude. Recently, Reames (2023) revisited this study adding examples that better illustrate the evolving role of the shock source as it propagates outward from the corona. A collection of sample time distributions of protons vs. longitude from this work is shown in **Figure 1**. While the SEP events in **Figure 1** are all different, we expect that slices of a single event at different longitudes could show similar features.

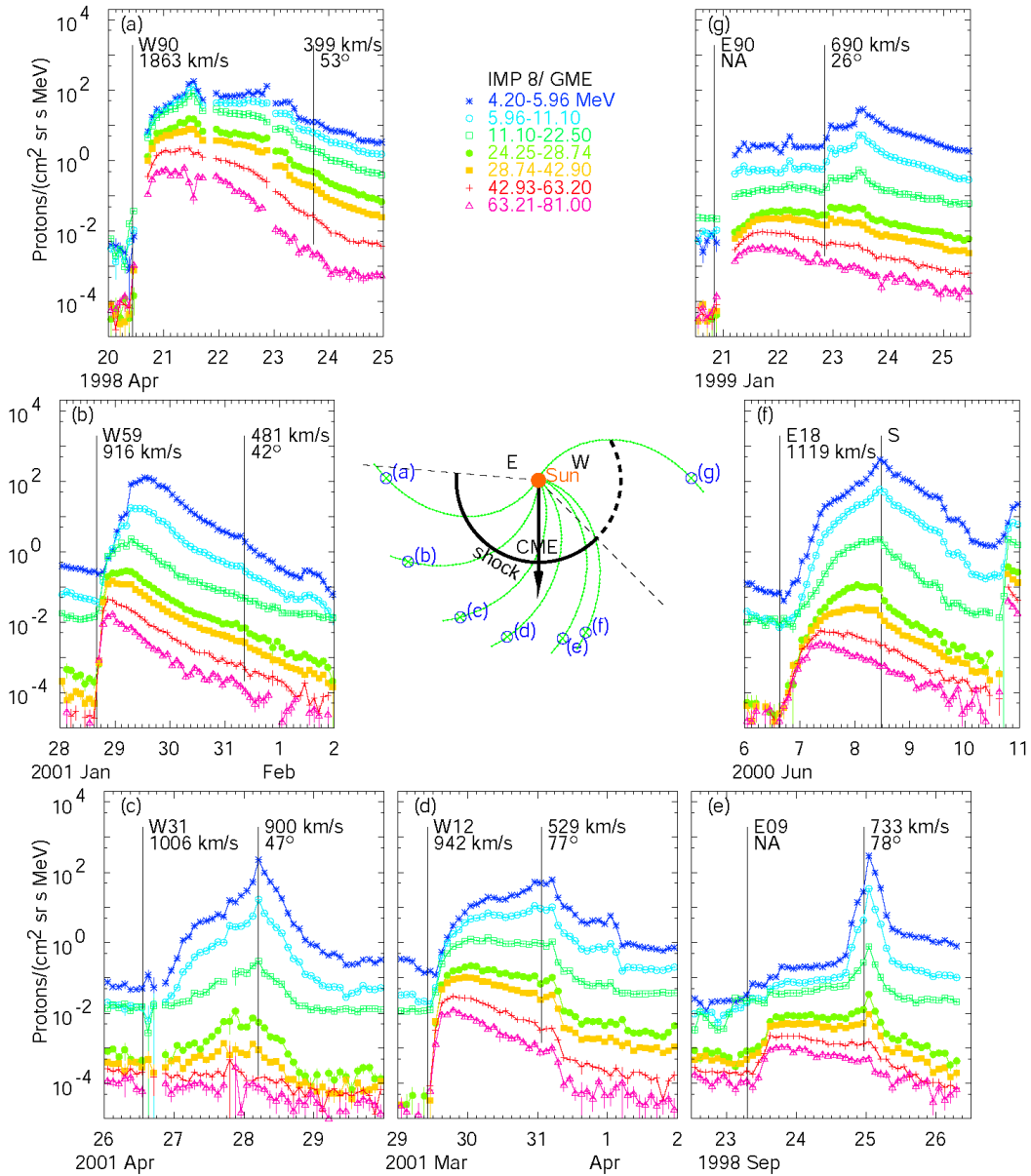

**FIGURE 1** IMP 8/GME proton intensities of the listed energies are shown vs. time in each panel (a) through (f) around a central map showing their nominal distribution around a fixed CME-driven shock source. Event onset times in each panel are flagged with the source longitude and the CME speed (when available) and the times of shock passage are noted with the shock speed and $\theta_{Bn}$, the angle between the field $B$ and the shock normal (when available). All panels have the same intensity scale. Dashed lines in the central map illustrate that a shock must first encounter any field line at its footpoint on the east flank, but may first strike it far from the Sun on the west (Reames 2023).

Historically, the intensity peaks at the shock seen on several of the profiles in **Figure 1** were called "energetic storm particle" (ESP) events. During their acceleration, particles are trapped near the shock by self-generated Alfvén waves (Bell 1978a, b; Lee 1983; 2005) creating an autonomous structure that can propagate onto new field lines where earlier accelerated particles may be absent, as in **Figure 1e** (Reames 2023).

## 3 Multispacecraft Observations

### 3.1 Pioneer

Some of the early *Pioneer* spacecraft were launched into Earth-like solar orbits, although coverage was only hours per day. When there was no spacecraft near Earth, this led to the awkward comparison of 15 MeV proton data from *Pioneer* 6 and 7 with GeV ground-level neutron-monitor measurements at Earth (Bukata et al. 1969). Fortunately, McKibben (1972) was able to include data from IMP 4 near Earth. While most of these spatial distributions were analyzed in terms of adjustable coefficients in the fashionable perpendicular diffusion, rather than shock acceleration, McKibben also noted that late in SEP events, proton intensities could be identical over large spans of longitude. Later named "reservoirs" by Roelof et al. (1992) these regions involved particles quasi-trapped magnetically behind the shock; as the volume of this magnetic bottle expands adiabatic deceleration (e.g. Kecskeméty et al. 2009) decreases all intensities, preserving spectral shapes (Reames et al. 1997; Reames 2013). Extreme SEP scattering, once used to explain this slow decay, is actually found to be negligible in reservoirs since particles from new $^3$He-rich events travel scatter-free across them (Mason et al. 1989; Reames 2021a).

### 3.2 Ulysses

Whenever there was a dependable spacecraft stationed near Earth, like IMP 8, any traveling spacecraft, such as the solar-polar-orbiting *Ulysses*, allowed two-spacecraft comparisons. The reservoir comparisons of Roelof et al. (1992) found uniform intensities behind the shock, extending radially over 2.5 AU from IMP 8 to *Ulysses*. In fact, Ulysses observed reservoirs at heliolatitudes up to >70⁰, N and S (Lario 2010), and in other electron observations (Daibog et al. 2003).

### 3.3 *Helios,* IMP, and *Voyager*

The two *Helios* spacecraft followed neighboring solar orbits from 0.3 to 1.0 AU beginning in 1974. Beeck et al. (1987) used data from these spacecraft to study the radial and energy dependence of diffusive scattering of protons, while, in a larger study, Lario et al. (2006) fit the peak intensities and the fluence of events to the form $R^{-n} exp[- k(\phi - \phi_0)^2]$ where *n* and *k* are constants, *R* is the observers radius in AU, $\phi$ is the solar longitude of the observer and $\phi_0$ is that of the event centroid. For a Gaussian distribution, $k = (2\sigma^2)^{-1}$. Lario et al. (2006, 2007) also considered the time variation of the point where the observer's field line intercepts the expanding shock source, an important feature of shock acceleration defined by Heras et al (1995).

*Helios* provided an excellent opportunity to study spatial distributions of shock-accelerated particles as well as their reservoirs, especially during 1978 when the *Voyager* spacecraft were nearby (Reames et al. 1996, 1997, 2013; Reames 2023). An especially interesting event is shown in **Figure 2** where the intensities of SEPs at *Voyager 2* **(Figure 2a or 2e)** do not begin to increase until after the shock passes S1 **(Figure 2b),** presumably because this is where it first intercepts the field line to *Voyager*. Proton intensities then slowly rise as *Voyager* becomes connected to stronger and stronger regions of the approaching shock until the ESP structure finally arrives at S4 on 6 January. Notice in **Figure 2a** that the peak intensities near the shock are similar at all four spacecraft. The ESP structure forms as protons streaming away from the shock generate resonate waves of wave number $k \approx B/\mu P$ where $B$ is the field strength, $P$ is the proton rigidity and $\mu$ is the cosine of its pitch angle (Lee 1983, 2005; Ng and Reames 2008; Reames 2023). These waves trap particles in ESP structures. As the structure moves out to lower $B$, the resonance shifts so that high energies preferentially leak away early, as also seen in **Figure 1e**. *Voyager* sees the pure "naked" ESP event with few of the streaming protons that created it. The shock that eventually arrives at *Voyager* generated the intense early streaming protons seen early by *Helios 1* and an intermediate structure as it passed *Helios 2* and IMP 8.

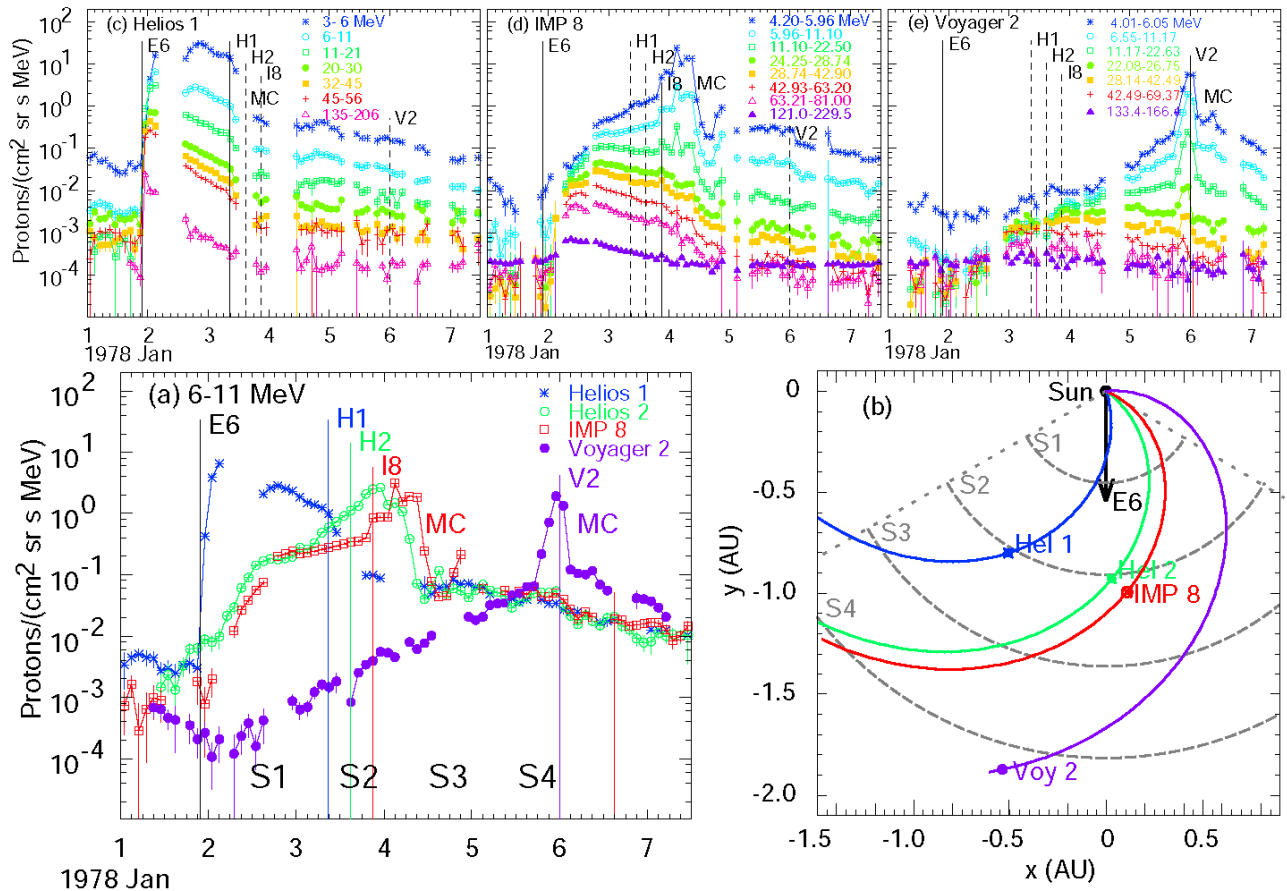

**FIGURE 2** In **(a)**, intensities of 6–11 MeV protons are compared for *Helios 1, Helios 2*, IMP 8, and *Voyager 2* during the 1 January 1978 SEP event, while **(b)** shows the spatial configuration of the spacecraft on their initial field lines and stages in the expansion of the CME-driven shock at S1, S2, etc. are sketched. Intensity-time profiles for a full list of energy intervals are shown for **(c)** *Helios 1*, **(d)** IMP 8, and **(e)** *Voyager 2*. MC is a magnetic cloud from the original CME (Burlaga et al. 1981). Onset time of the event is flagged by E6 and shock passage at each spacecraft by H1, H2, I8, and V2 (Reames 2023).

The spacecraft distribution for the event shown in **Figure 2** is unique, in that spacecraft sample the SEPs produced as the source shock moves radially. The width of this CME is limited, but the spacecraft are positioned to follow the evolution of the event: (1) well-connected *Helios 1* samples its central production near the Sun, (2) *Helios 2* and IMP 8 sample its production near 1 AU, and (3) *Voyager 2* scans its western flank then samples its central strength as it passes 2 AU. A fortuitous occurrence: NASA did not design *Voyager* as a complement to *Helios*.

Another fortuitous observation with these spacecraft occurred in September 1978 and is shown in **Figure 3**. In **Figure 3a**, well-connected IMP 8 shows a fast rise in intensities and a peak near the time of shock passage, while *Helios 1* and *2*, far around the west flank, rise more slowly then join IMP 8 in a reservoir behind the shock on 25 September. Distant *Voyagers* show a slow SEP rise to a plateau near the time S2 where IMP 8 joins the same intensity once it is no longer constrained by the east flank of the shock. Then at S3, the *western* flank of the shock strikes field lines that send particles sunward to IMP 8 and outward to *Voyager 2*, then to *Voyager 1*. This second SEP peak is clearly seen at energies above 40 MeV in **Figures 3c** and probably even in **3d**. A "left" and then a "right" from the same shock; at *Voyager* the two peaks are comparable in size.

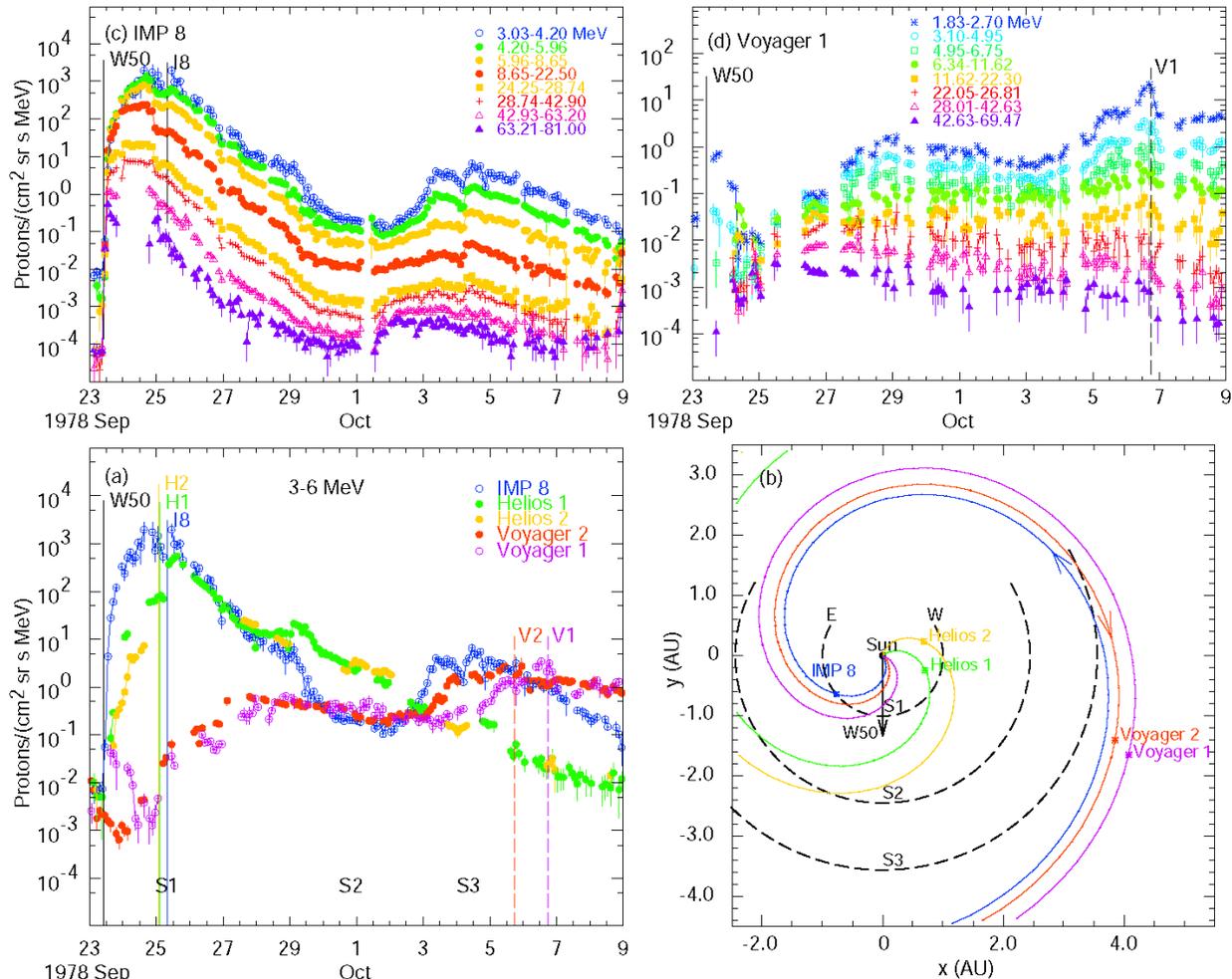

**FIGURE 3** In **(a)** intensities of 3–6 MeV protons are compared for IMP 8 (**blue**), *Helios 1* (**green**), *Helios 2* (**yellow**), *Voyager 1* (**red**), and *Voyager 2* (**violet**) in the 23 September 1978 SEP event, while **(b)** maps the configuration of the spacecraft on their initial field lines and shows the expansion of a CME-driven shock at S1, S2, and S3. Onset time of the event is flagged as W50 and shock passage at each spacecraft as H1, H2, I8, V1, and V2. In **(b)**, the **western** flank of the shock S3 intercepts the **blue** and **red** fields, where arrows direct particles accelerated sunward to IMP 8, then outward to *Voyager 2*, respectively, and later to *Voyager 1*. Intensity-time profiles for a full lists of energy intervals are shown for **(c)** IMP 8, and **(d)** *Voyager 1* (Reames 2023).

Note that the second peak in the IMP 8 data in **Figure 3c** shows significant velocity dispersion corresponding to the ~6-AU path inward from the new source (actually a larger delay than in the first peak) while none is seen at Voyager 2 (**Figure 3d**) near this source. Also the intensities at IMP 8 and Voyager are quite similar in the second peak; any new injection from the Sun would have produced a huge difference in intensities like that seen in the first peak because of the great difference in radial distances since IMP 8 and the Voyagers seem to be on similar field lines. At the second peak, the shock has filled these field lines, forming a reservoir with similar intensities at IMP and Voyager.

Another parameter that can depend upon solar longitude is the solar particle release (SPR) time derived from velocity dispersion (Tylka et al. 2003; Reames 2009a, 2009b). If the first particles of each energy to arrive at the spacecraft have been released at nearly a single time, the SPR time, and have scattered little, their travel time will be $dt = L/v$ where L is the field line length from source to observer and $v$ in the particle velocity. Plotting the onset times vs. $v^{-1}$ will yield a linear fit with slope L and intercept at the SPR time. **Figure 4b** shows such a plot from Reames and Lal (2010) for the spacecraft distributed as shown in **Figure 4a** during the GLE of 22 November 1977. The parabolic fit for the height in **Figure 4d**, not from this event, is a fit of 26 GLEs observed from Earth (Reames 2009b); presumably it is a first-order correction for weaker, more slowly evolving, shock flanks. Gopalswamy et al. (2013) fit the SPR height for GLEs at Earth directly to the source longitude (uncorrected for foot-point motion). Type II bursts begin near ≈1.3 $R_S$ but they certainly need not correspond to the same field line longitude or shock physics as the measured SEPs. Non-relativistic electrons that produce type II bursts do not resonate with Alfvén waves like ions and hence are more likely to be accelerated by quasi-perpendicular regions of the shocks.

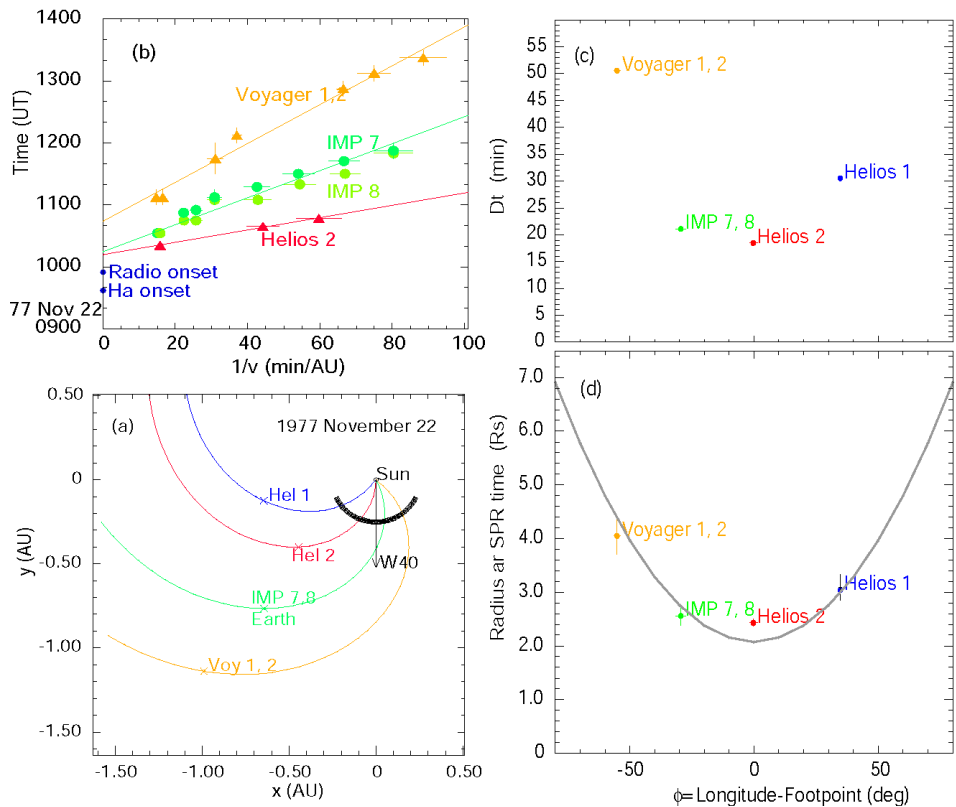

**FIGURE 4** (**a**) shows the distribution of spacecraft during the 22 November 1977 SEP event at W40, (**b**) shows plots and least--square fits of the particle onset times vs. $v^{-1}$ at available energies for several spacecraft, (**c**) shows the time delay and (**d**) shows the corresponding shock height of the SPR time vs. solar source longitude minus the footpoint longitude. Data from *Helios 1* are limited (Reames and Lal 2010). The parabola in (**d**), from Reames (2009b), may just be a first order correction for the showing of the shock on its flanks.

Thus *Helios* allowed spatial comparisons of the beginnings of events, their evolution and ESP events at the middle, and their reservoirs at the end (Reames et al. 1997), noted earlier.

### 3.4   STEREO, with *Wind*, ACE, and SOHO

The launch of *Wind* in November 1994 began a new era in SEP coverage from Earth and it was later joined by SOHO and ACE adding different capabilities. This coverage still exists in 2023. STEREO ahead (A) and behind (B) were launched together in late 2006 at the beginning of solar minimum. By September 2012 they had reached ±120º longitude, providing three equally-spaced observing points around the Sun, optimal coverage for finding the most extensive SEP events.

With STEREO it has become even clearer that SEP events can be quite extensive. For > 25 MeV proton events, Richardson et al. (2014) found 17% spanned three spacecraft, 34% two spacecraft, 36% one spacecraft, with 13% unclear. Studying the abundances of H, He, O, and Fe at 0.3, 1, and 10 MeV amu$^{-1}$, Cohen et al. (2017) found only 10 three-spacecraft events, out of 41.

Most of the studies of particle distributions with STEREO have involved Gaussian fits of the intensity peaks or fluences at three longitudes. Of course, three points determine a parabola and a Gaussian is a parabola in logarithmic space, so Gaussians tend to fit the data very well. Xie et al. (2019) studied 19 – 30 MeV protons in 28 events finding $\sigma = 39° \pm 6.8°$. Paassilta et al. (2018) compiled a list of 46 wide-longitude events above 55 MeV, of which seven were suitable, averaging $\sigma = 43.6 \pm 8.3°$ and for 14 events with E > 80 MeV, de Nolfo et al. (2019) found an average $\sigma \approx 41°$. Lario et al. (2013) studied the 15–40 and 25–53 MeV proton peak intensities and found $\sigma \approx 45 \pm 2°$ for both proton energy ranges. Cohen et al. (2017) found average three-spacecraft events centered at `$\approx 22 \pm 4°$ west of the flare site and $\approx 43 \pm 1°$ wide; they found no dependence of the width on the charge-to-mass ratio $Q/A$ of the elements. This lack of dependence on $A/Q$ seems to argue against lateral diffusive transport. Kahler et al. (2023) fit the data of 20 MeV protons to hourly Gaussians, showing substantial broadening of the distribution with time, especially on the western flank where the shock expanded across new spiral field lines. However, all of these similar widths and parameters only apply to the largest $\approx 20\%$ of events that span three spacecraft. We have no widths from the one-spacecraft third of the events. For electron events, Klassen et al. (2016) examined events for closely spaced (<72°) STEREO spacecraft and found they were not well fit by Gaussians while Dresing et al. (2018) pointed out that peak intensities used for the Gaussian fits may not represent the real spatio-temporal intensity distribution as the intensity peaks may have been measured at different times.

Some of the space-time coupling in SEP distributions is illustrated by the event shown in **Figure 5**. Here the shock itself is seen at each spacecraft, yet the slight onset delay at *Wind* suggests it misses the base of that field line. Both STEREO spacecraft show fast intensity increases, followed by early declines, but the intensity at *Wind* peaks well after the shock passage. How do we distinguish variations of space and time? When does a Gaussian spatial distribution apply? Is the Gaussian formed by the peaks, or by the fluences; is it spatial or also temporal? Even hourly Gaussians miss the true behavior late in an event.

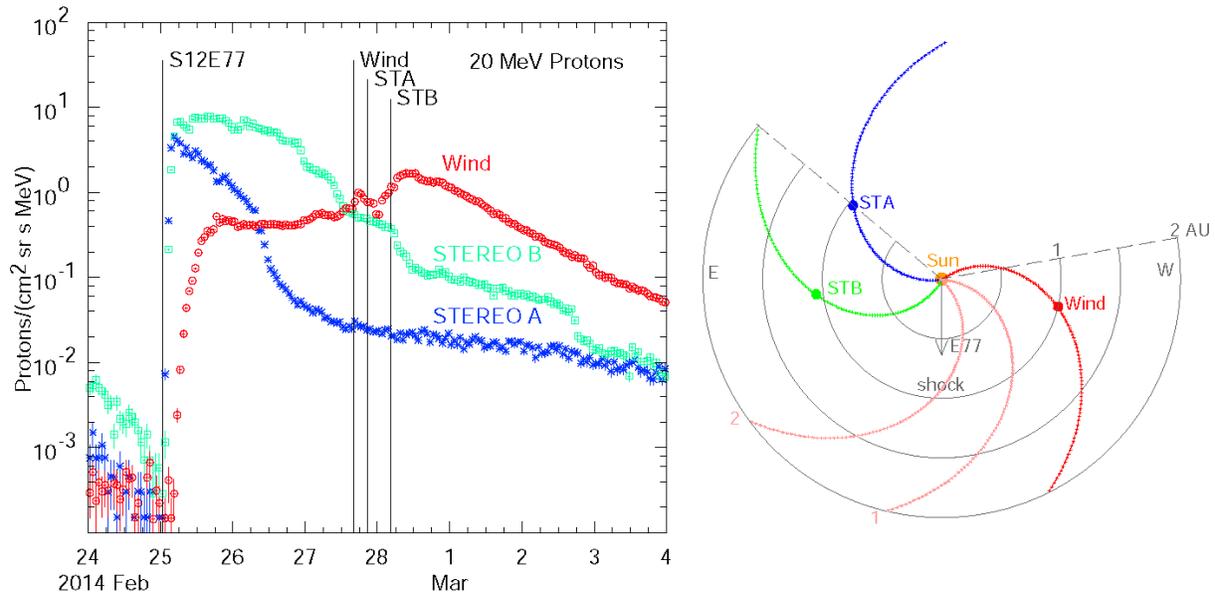

**FIGURE 5** shows time variations of ~20 MeV protons in the large three-spacecraft SEP event of 25 February 2014 on the **left** and the map of spacecraft configuration on the **right**. Vertical time flags mark the event onset from S12E77 and the shock passage times at each spacecraft. In the map, circles follow expansion of a spherical shock wave, and initial Parker spirals connect the Sun with each spacecraft; these field lines would be distorted as the shock passes. Pink field lines measure solar rotation (see text).

The two pink field lines labeled 1 and 2 in the map in **Figure 5** will be carried, from their initial positions shown, to the red location of *Wind* by the time the shock reaches 1 or 2 solar radii, respectively, because of solar rotation. Hence the time profile we see at *Wind* is greatly enhanced behind the shock by SEPs swept in on field lines that were much more centrally located initially. Meanwhile, STEREO A is rapidly acquiring field lines with decreased intensities of SEPs that have rotated from behind the Sun, and STEREO B eventually joins it in a suppressed reservoir region late on 2 March. Can SEP models follow this mixture of space-time variations? Time variations longer than about a day need to accommodate the 13⁰ day$^{-1}$ solar rotation. This event evolves over a week.

Note also in **Figure 5** that the accelerating shock wave is seen at all three spacecraft, as is often the case. These shock waves are able to wrap around the Sun, crossing field lines that the particles alone cannot cross. Thus STEREO shows us the way the physics of shock acceleration can easily replace the early confusion of the "birdcage model," "coronal diffusion," and the "fast propagation region" – diffusion from a point source – that once held sway.

### 3.5   SEP Element Abundances

Spatial distributions of fluences of element abundances were studied extensively by Cohen et al. (2017) using STEREO and ACE data. The Gaussian distributions of H, He, O and Fe were found to be similar in ten three-spacecraft events, suggesting that the widths are independent of rigidity or transport. They found the average three-spacecraft Gaussian distributions to be $43 \pm 1°$ wide, although with significant variations. None of these large three-spacecraft SEP events showed any of the enhancements of heavy elements, e.g. Fe/O, typically found in shock-reaccelerated impulsive suprathermal ions of SEP3 events. An analysis of the *A/Q*-dependence of the element abundances in a typical large SEP4 event is shown in **Figure 6**, where power-law fits of flat or suppressed heavy elements extend to include protons (Reames 2020, 2022b).

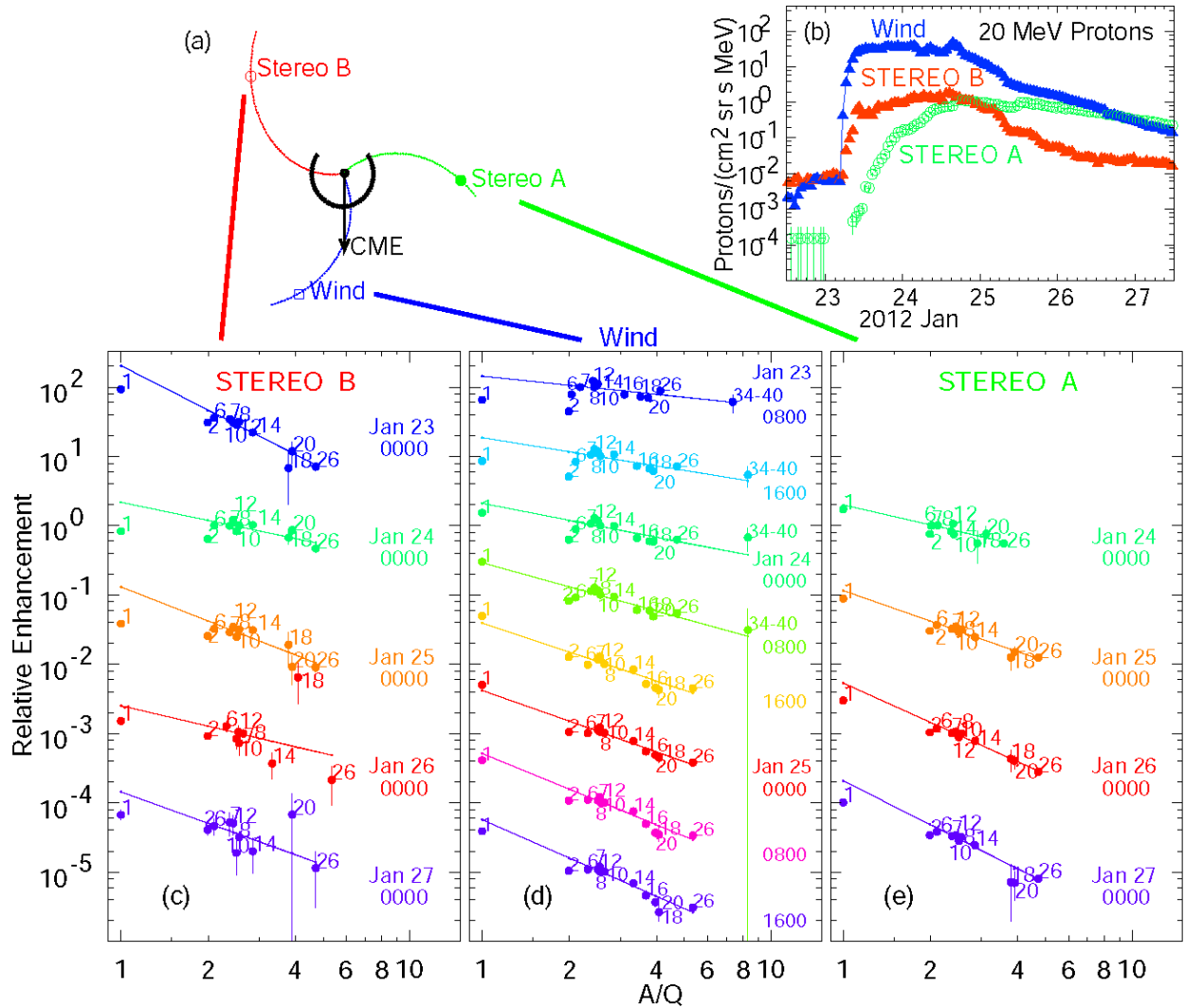

**FIGURE 6** (**a**) shows a map of the spacecraft distributions during the SEP event of 23 January, 2012; (**b**) shows the corresponding intensity-time profiles of 20 MeV protons at each spacecraft. The abundance enhancements, relative to SEP-average (coronal) abundances, for elements noted by $Z$, are shown vs. $A/Q$ for the listed time intervals for (**c**) STEREO B, (**d**) *Wind*, and (**e**) STEREO A. Best fit power laws vs. $A/Q$ for elements with $Z > 2$ are shown extended down to protons at $A/Q = 1$.

For each time period, the observed abundances of $Z \geq 6$ ions are divided by the corresponding coronal abundances and fit to power-laws. Most SEP4 events show flat fits, i.e. abundances the same as coronal, or declining power laws as in Figure 6. The declining power laws may result from reduced scattering of heavier ions that allows them to leak more easily from the acceleration region SEP3 events have both enhanced protons and power-law heavy-element enhancements that rise with $A/Q$, reflecting shock acceleration of seed populations of normal ambient coronal ions as well as impulsive suprathermal ions with their characteristic high-$Z$ enhancement. SEP3 events can be very large. In solar cycle 23, about half of the GLEs were SEP3 events (Reames 2022a) and half SEP4 events. In the weaker cycle 24, when STEREO was available, we find only one of the two-spacecraft events listed by Cohen et al. (2017) that was an SEP3 event (4 August 2011); this is an SEP3 event at *Wind* (Figure 10 in Reames 2020) and shows similar enhancements at STEREO A, although with poorer statistics. Why are there so few wide SEP3 events? Is it the weak solar cycle or are SEP3 events inherently narrow, perhaps because the primary pools of seed particles are confined?

### 3.6 Impulsive SEP Events

A distinction of impulsive SEP events is that their longitude spread is much more limited than that of gradual events (Reames 1999) but the width of that distribution depends upon the sensitivity of the instruments observing them and has increased somewhat with time (e.g. Reames et al. 2014a). A significant factor in this width is that variations of the solar wind speed vary the longitude of the footpoint of the observer's field line, but there is also variation due to the random walk of the footpoints of the field lines (Jokipii and Parker 1969; Li and Bian 2023) prior to the event.

A search for $^3$He-rich events between ISEE 3 and *Helios* (Reames et al. 1991) found several corresponding events. The well-studied event of 17 May 1979 (e.g. Reames et al. 1985) with $^3$He/$^4$He ≥ 10 was seen with similar $^3$He enhancement by *Helios 1* near 0.3 AU as shown in **Figure 7** as a sharp spike of very short (~1 hr) duration presumably resulting from reduced scattering. Associated electron trajectories were tracked spatially, using the direction and frequency of the radio type III burst as measured from ISEE 3 (Reames et al. 1991).

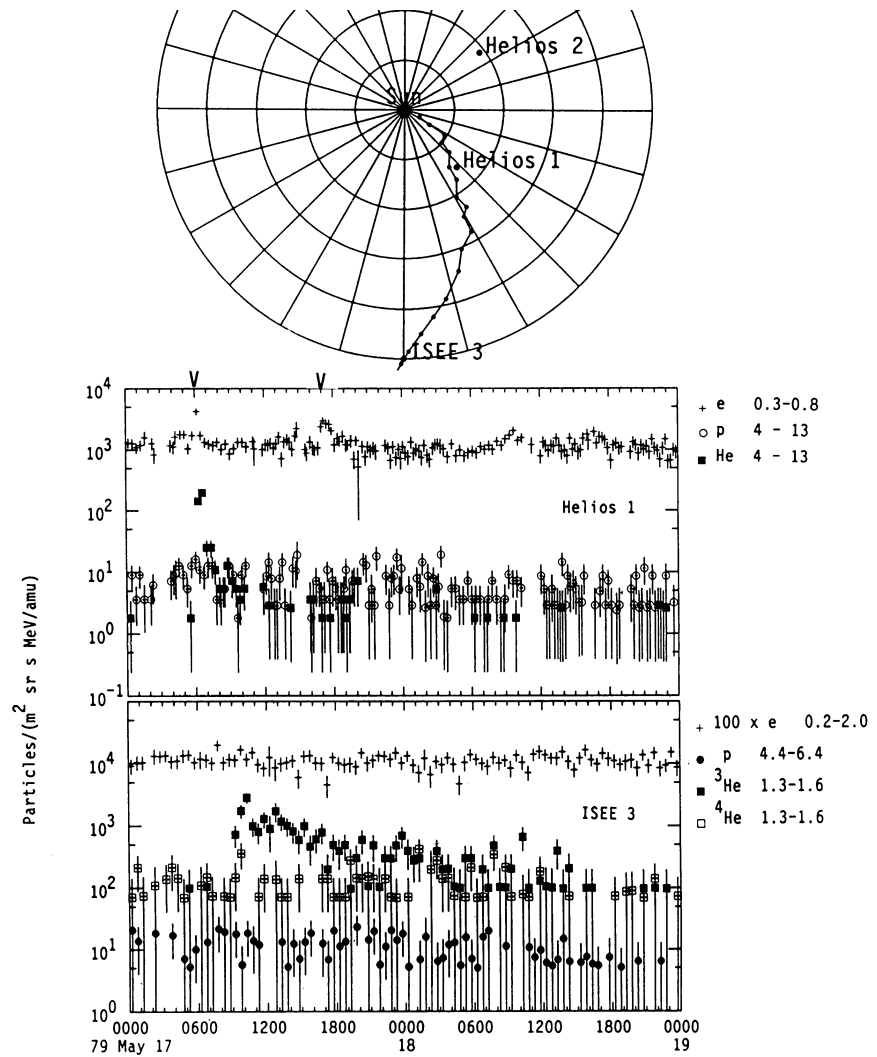

**Figure 7** The extremely ³He-rich event of 17 May 1979 is shown by the track of its radio type III burst in the **upper** panel, and time histories of electrons and ions indicated at *Helios 1* and ISEE 3 in the panels below. Indicated energies are in MeV for electrons and protons, MeV amu$^{-1}$ for He. Arrows above the central panel mark two times of event sources. (Reames et al. 1991).

The time history for the Kiel Instrument on *Helios 1* in **Figure 7** shows composite He; isotope ratios were tabulated separately using pulse-height data. Despite differences in He energies at the two spacecraft, both show the event at 0550 UT with ³He/⁴He ≈ 10 and that at 1700 UT with ³He/⁴He ≈ 1.

In the early STEREO era, before the spacecraft were widely separated, Wiedenbeck et al. (2013) found electrons and ions ±20° ahead and behind Earth, during an event that occurred during a solar quiet time. The ions showed strong intensity gradients. The event was associated with a weak CME. Also, Klassen et al. (2015) studied an electron beam event early in the STEREO mission.

The discovery of fast CMEs associated with impulsive SEP events (Kahler et al. 2001) suggested that impulsive SEPs could be spread laterally when nearly radial shocks in these SEP2 events distributed particles across spiral field lines. Particles from SEP1 jets without fast shocks would be expected to follow any open field lines from the jet, presumably less widely distributed. Suggestions of greater spreads for SEP2 impulsive events with shocks, based upon current understanding, have not been explored, mainly because the available spacecraft are too widely separated.

Opportunities for multiple measurements of impulsive SEP events are rare. Perhaps the best opportunity is for measurements of the radial variation in SEP scattering between *Parker Solar Probe* (PSP) or *Solar Orbiter* and spacecraft near Earth. Are the time profiles of SEPs at PSP near the Sun similar to those of X-ray profiles of the event? To what extent are the jets that release SEPs part of more-extensive flaring systems? Magnetic reconnection that that opens some field lines must also close others (e.g. see Figure 4 in Reames 2021b), but reconnecting closed field lines with other closed lines would allow no escape.

## 4    Discussion

STEREO has given us an improved sense of the widths of some of the most extensive SEP events, but are the SEP distributions really characterized by nice smooth Gaussians? Are the SPR times or heights really parabolic? During the *Helios* era, when the *Voyagers* happened by, a few events showed us greater complexity (e.g. **Figures 2** and **3**). Such events allow us to explore the underlying physics. STEREO also allows 3D modeling of the CME and its shock (e.g. Rouillard et al. 2011, 2012; Kouloumvakos et al. 2019; Zhang et al. 2023) but this is only beginning to be extended to a 3D modeling of the SEP distribution this shock produces. Correlations between SEP peaks and shock properties at the base of the observer's field line (e.g. Koulomvakos et al 2019) may have reached their limits. SEP peak intensities and times are controlled, and often limited, by transport and are not determined by any single point on the evolving shock; transport and shock contributions both vary with time in complex ways. No single parameter represents an entire SEP event very well.

Kahler's (1982) "big flare syndrome", while expressed for a specific case, should be a general warning that correlations do not imply causality. Increasingly energetic magnetic reconnection events at the Sun can spawn bigger flares, faster CMEs, and larger SEP events. They are all correlated, yet Hα flares do not *cause* CMEs, or GLEs; ultimately they are all consequences of magnetic reconnection. Later, Kahler (1992) asks "how did we form such a fundamentally incorrect view of the effects of flares after so much observational and theoretical work?" We need clearer resolution of the underlying physics that does connect a cause and its effects.

Gopalswamy et al. (2013) sought to understand possible differences between the GLE of 17 May 2012 and six other large SEP events with similar CME speeds that were not GLEs. This event had a smaller flare (M5.1) than any of the others. A large effect was the small latitudinal distance of the shock-nose from the ecliptic, i.e. the GLE was better connected to Earth. The GeV protons in this GLE were all produced within ~8 minutes. This suggests a highly localized region of production might occur soon after shock formation (Ng and Reames 2008) and may differ from the global properties of the CME, discounting correlations. Incidentally, the 17 May 2012 event was the only GLE included in the study by Cohen et al. (2017) and it had too few heavy ions to be studied at three spacecraft; it was not one of the ten three-spacecraft SEP events.

It has also been observed that GLEs are more likely when shocks pass through solar streamers where the higher densities and lower Alfvén speeds produce higher Alfvénic Mach numbers (Liu et al. 2023) and regions of higher $\theta_{Bn}$ (e.g. Kong et al. 2017, 2019) that can enhance acceleration.

Intensities of GeV protons only exceed GCRs for short periods and are too weak to be a practical radiation hazards. However, ~100 MeV protons are much more numerous and persistent and hence a significant hazard to astronauts outside the Earth's magnetic fields. When an event is near central meridian, intensities of high-energy SEPs can be extended and increased by the ESP event when the nose of the shock passes over us. Some historical examples are shown in **Figure 8**. For western sources we see the greatest effect of the shock nose early, but the shock then weakens toward our longitude and weakens in strength out to 1 AU. The early SEP intensities at 1 AU are constrained by wave growth which can establish the "streaming limit" early (Reames and Ng 1998, 2010, 2014; Ng et al. 1999, 2003, 2012) while ESP intensities are unbounded. Thus, the spatial distribution of SEP intensities at the shock, i.e. the ESP event, is of both fundamental and practical importance. The SEP events of October 1989 shown in **Figure 8** are the basis of SEP "storm shelter" radiation shielding requirements for astronauts in missions beyond low Earth orbit (Townsend et al. 2018).

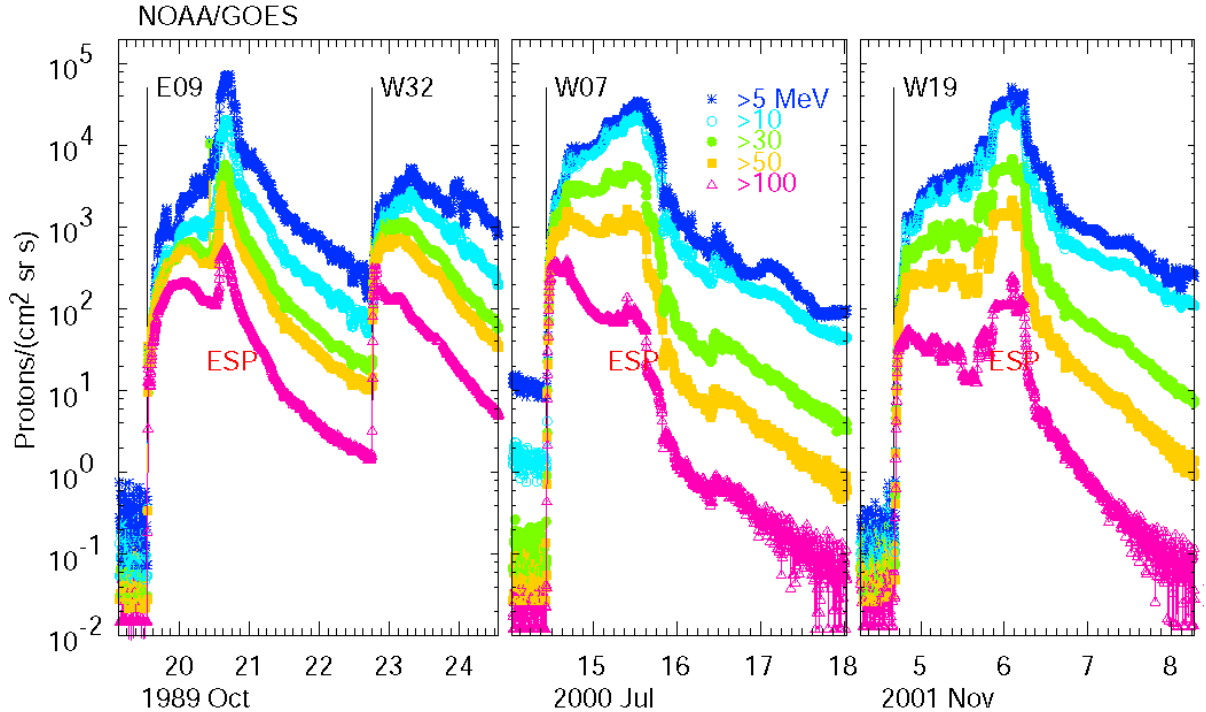

**FIGURE 8** High-energy proton intensities during historic large SEP events near central meridian can be enhanced and extended in time by their ESP events, even at energies >100 MeV (a significant radiation hazard). Early intensities are bounded by the "streaming limit" but the ESP peak is not. All of these events are GLEs which often occur near central meridian, allowing the ESP to surface.

The ESP structure is formed very early in an event when protons streaming away from the shock amplify resonant Alfvén waves (Stix 1992; Melrose 1980). Waves trap ions of a given rigidity, scattering them back and forth across the shock so they gain velocity on each transit, subsequently amplifying waves of lower $k$ (longer wavelength) which trap ions of higher $P$, etc., creating the ESP structure of energetic ions trapped around the shock (Lee 1983, 2005). At the streaming limit, higher SEP intensities simply grow more waves, trapping more ions back near the shock. In gradual SEP events, the ESP structure always exists. When the shock is near the Sun, the ESP is initially hidden among the same SEPs that form it as they stream away. Of course, the shock we see at 1 AU was much stronger when it began near the Sun. The shock and the ESP event can explain *all* of the SEPs in the gradual event, early and late. It is a key to the physics of SEP acceleration in these events, whether it emerges at your particular longitude or not. The basic ESP structure will only be exposed at some longitudes in some events.

When the flank of a shock crosses to new field lines, where the early emission is absent, the "naked" ESP event emerges as seen at *Voyager* in **Figure 2e** or at IMP 8 in **Figure 1e**. The ESP structure persists as the shock moves outward, but the CME speed decreases and decreasing $B$ shifts the resonance so the highest energy ions preferentially begin to leak away. Earlier peaks of ~30 – 100 MeV protons are seen along with the naked ESP events in these figures because they leaked out *after* the ESP events have crossed to the new field lines.

Generally, the time of the peak intensity is a contest: the initial intensities are bounded at the streaming limit while the ESP peak is unbounded, yet the highest energies in the ESP begin to leak away as the shock slows and moves out to lower $B$. Understanding the high-energy particle acceleration of SEPs is about understanding shock acceleration, wave trapping, and the formation, spatial extent, and persistence of ESP events. The large events in **Figure 8** are relatively rare, and we tend to assume that the same physics can be studied at lower energies. However, the physics changes for those energies where resonant waves dominate – a function of space and time. A few general questions do arise:

1) Comparing **Figure 1c** and **Figure 1e**, how important is $\theta_{Bn}$ in maintaining the ESP event?

2) The increasing intensity ramp at *Voyager* in **Figure 2a** involves SEPs leaked from the approaching ESP event. Does it show strengthening of the shock with longitude toward the east or mainly increasing proximity?

3) Are the SEP3 events with enhanced heavy ions limited in longitude? Surely seed-particle populations in an event can change with longitude; where are the events that change from SEP3 to SEP4? Why are STEREO three-spacecraft events all SEP4 events? Half of GLEs are SEP3s.

4) How does the longitude extent of the SEPs compare with that of the shock itself? How does each vary with time?

5) Are GLEs or other high-energy events more limited in longitude? Would they be one- or two-spacecraft events for STEREO? Are broader shocks weaker or less efficient?

We should expect that shocks can accelerate whatever seed population or populations they encounter, wherever they go. We cannot exclude mixtures; we can only distinguish which one dominates at high *Z* and thus label an event SEP3 or SEP4.

We are constantly hampered by the correlated mixture of space and time. The footpoint of a field line from Earth lies 50 - 60⁰ to our west. As our connection point on the shock scans to the east, that shock also weakens with time. How much is change with longitude; how much is change with time? The only resolution is to measure spatial variations with a scale substantially less than 50⁰. By the time the shock arrives at 1 AU, it has mixed SEPs from ≈50⁰ of solar longitude; in longer time intervals solar rotation causes greater mixing.

## 5   What Is Needed Next?

SEP evolution in space and time is complicated. STEREO spacecraft separation from Earth of ~120⁰ was much too coarse to resolve the SEP-shock evolution we happened to observe in a *Helios*-IMP-*Voyager* period. What kind of observations would help? Multiple points with better spatial resolution. Consider two primary spacecraft, equipped as STEREO was, with coronagraphs and in situ instruments, which are fixed in Earth's solar orbit at ±60⁰ from Earth as shown in **Figure 9**. On each side, between each primary spacecraft and Earth are two much smaller spin-stabilized spacecraft (similar to *Wind*), 20⁰ apart, each capable of measuring SEPs, magnetic fields, and solar-wind plasma to map SEP events, shocks, and interplanetary CME structures. This configuration presumes that capabilities similar to a primary spacecraft are preexisting near Earth.

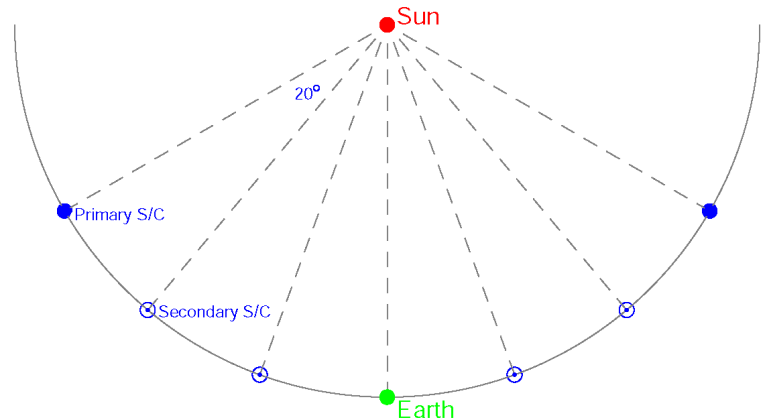

**FIGURE 9** A possible multi-spacecraft configuration of the future might include two primary spacecraft, similar to STEREO, and four small spacecraft measuring SEPs, plasma, and fields, to study in situ spatial distributions of SEPs, interplanetary CMEs, and shock waves. The spacecraft would maintain their positions relative to Earth for years, spanning at least one solar maximum.

The ≈20⁰ spacing would provide at least two measurements between the longitude of a spacecraft and that of its typical solar magnetic footpoint. It would allow meaningful coverage of many spatially small- and moderate-sized events with sampling of variations along the shock for the larger ones. "Smaller" SEPs does *not* mean weaker. Of course, it would be better to have more-complete coverage, but the configuration in **Figure 9** represents a major improvement in spatial resolution, and reasonable (33%) coverage, at a modest increase in cost and complexity over the dual-spacecraft STEREO mission. A spacing of ≤ 20⁰ seems essential and provides a reasonable tradeoff between spacing and coverage. This would provide a map of the shock strength, direction of propagation, and $\theta_{Bn}$ at up to seven points at 1 AU that could be compared with the coronagraph mapping near the Sun, and could give seven SEP/ESP profiles with differing onset times and intensities. What does the shock front really look like? Is the nose of the shock a hot spot that could produce a localized GLE; for what longitude does the peak intensity arrive with the shock?

Are the highest energies in SEP events limited to short time periods and small spatial intervals as Gopalswamy et al. (2013) conclude? If so, what physics defines those intervals and why do they differ from the apparent STEREO finding of similarly broad Gaussians for most energy bands? Up to some energy, we would expect high SEP intensities to generate enough waves to extend the spatial trapping and the duration of an event (e.g. **Figure 8**), although not at the absolute peak energy where there are yet few resonant waves. These peaks could be highly localized in well-connected regions that become poorly correlated with the average properties of the CME. Also, in some events, particle trapping is increased by the presence of multiple shock waves as in the 14 July 2000 "Bastille Day" GLE (Lepping et al. 2001). To what extent can the SEP energy profile and the peak energy be predicted from an early coronagraph map of the shock strength? We would have an extra day to predict the strength of the ESP event.

## 6    Conflict of Interest

The author declares that this research was conducted in the absence of any commercial or financial relationships that could be construed as a potential conflict of interest.

## 7    Author Contributions

## 8    Funding

No institutional funding was provided for this work.

## 9 Acknowledgments

The author thanks Ed Cliver for helpful discussions.

# Solar Energetic Particles on Multispacecraft Missions

D V Reames

Solar Energetic Particles on Multispacecraft Missions